# Acoustic-electroelastic modeling of piezoelectric disks in high-intensity focused ultrasound power transfer systems

**Aarushi Bhargava[a] and Shima Shahab[a, b,1]**

[a] Department of Biomedical Engineering and Mechanics, Virginia Polytechnic Institute and State University, Blacksburg, VA 24061, USA

[b] Department of Mechanical Engineering, Virginia Polytechnic Institute and State University, Blacksburg, VA 24061, USA

**Abstract**

Contactless ultrasound power transfer (UPT) has emerged as one of the promising techniques for wireless power transfer. Physical processes supporting UPT include the vibrations at a transmitting/acoustic source element, acoustic wave propagation, piezoelectric transduction of elastic vibrations at a receiving element, and acoustic-structure interactions at the surfaces of the transmitting and receiving elements. A novel mechanism using a high-intensity focused ultrasound (HIFU) transmitter is proposed for enhanced power transfer in UPT systems. The HIFU source is used for actuating a finite-size piezoelectric disk receiver. The underlying physics of the proposed system includes the coupling of the nonlinear acoustic field with structural responses of the receiver, which leads to spatial resonances and the appearance of higher harmonics during wave propagation in a medium. Acoustic nonlinearity due to wave kinematics in the HIFU-UPT system is modeled by taking into account the effects of diffraction, absorption, and nonlinearity in the medium. Experimentally-validated acoustic-structure interaction formulation is employed in a finite element based multiphysics model. The results show that the HIFU high-level excitation can cause disproportionately large responses in the piezoelectric receiver if the frequency components in the nonlinear acoustic field coincide with the resonant frequencies of the receiver.

[1] Corresponding author: sshahab@vt.edu

## 1. Introduction

Ultrasound power transfer (UPT) has emerged as one of the more promising of all the other practiced techniques, namely inductive, capacitive, and microwave-based techniques, to wirelessly transfer power [1]. The underlying mechanism of UPT involves vibration-induced acoustic wave propagation from a piezoelectric transmitter that generates an elastic vibration-induced electrical response in a piezoelectric receiving element. The preference for UPT over other methods is due to use of acoustic waves at shorter wavelengths enabling the use of smaller sized receiver/transmitter, lower attenuation, higher penetration depth, no electromagnetic interference/losses, high directionality, and biological safety [1-4]. In recent years, UPT has found increasing applications in biomedical technology [5-8], data delivery [9, 10], and through-wall transmission [11-13]. Recent studies have explored the use of UPT to supply low electrical power (e.g. $1\mu W - 10 mW$ [14-16]) to biomedical implants to eliminate battery replacement issues and reduce the risks/maintenance costs for devices in inaccessible areas. For example, Cochran *et al*. [8] excited piezoelectric elements embedded in a fixation plate to provide current to electrodes placed at fracture sites to promote bone healing. Shi *et al*. [17] developed a MEMS-based piezoelectric ultrasonic energy harvester (PUEH) to power implants inside the body. By adjusting the frequency of the PUEH, they aimed to minimize the standing wave effect which can reduce the efficiency of the UPT system. Ozeri *et al.* [6, 7] demonstrated ultrasonic transcutaneous energy transfer from plane disc transducers to power implants in a pig muscle tissue up to 50 mm depth. Coupling of the low-power receivers with UPT and backscatter communication has also been used as a tool to build reliable neural recording systems [18].

In UPT, the power transfer efficiency is sensitive to the orientation of the receiver [19], depth of the transducer [20, 21], and acoustical scattering from the receiver [22]. Various configurations can be considered for UPT, as summarized by Shahab *et al.* [21]. These configurations include excitation of an array of receivers by a spherical source in the same domain [23], excitation of a receiver in a separate domain (e.g., as in transcutaneous UPT [6]), and enhanced power transfer by focusing of the source transmitted energy using high-intensity focused ultrasound (HIFU) transducer. The focusing of acoustic waves can also be achieved using passive acoustic holograms (lenses) to generate a multifocal pressure pattern [24, 25]. In this work, for the first time, a HIFU source is used for actuating a finite-size piezoelectric disk, i.e., the diameter-to-thickness ratio is higher than 0.1 and less than 20, in a UPT system.

HIFU has been used in the biomedical field for several years in various applications, such as drug delivery, therapeutic applications, and neurostimulation of muscles [26-30]. The two main advantages which support the usage of HIFU in UPT are: (1) its capability of focusing acoustic energy in a tight spot resulting in a localized, selective, and controlled actuation; and (2) the increased pressure obtained at the focal spot as compared to spherical or planar waves, thus increasing the efficiency of the system [31]. Consequently, the use of HIFU in UPT systems will enable a significant increase in power transfer efficiency as well as target the energy transfer to the desired receiver. The localization ability of HIFU also provides flexibility to use small-sized receivers for UPT applications. This energy concentration ability is particularly required in the powering of devices placed in sensitive environments. For such cases, the thermal effects associated with ultrasound are undesirable in the surrounding areas, such as the wireless powering of small-size implants in bodies or neural dust motes in the brain [18]. Our previous work of actuating polymer-based drug delivery containers inside the body, also demonstrates that the use of HIFU can achieve targeted response by localizing the thermal effect inside the polymers while maintaining the acoustic intensity in surrounding tissues below the FDA prescribed levels [32, 33].

The limited existing models on UPT cannot be applied to predict acoustic pressure fields from HIFU, as they do not model focusing of waves and its associated effects, such as diffraction and nonlinearity, in the HIFU-UPT system [1, 6, 20, 21, 24, 34]. Moreover, detailed and systematic investigations on understanding the physics of each of the individual entities involved in UPT are limited and assume linear acoustic wave propagation [35]. Shahab *et al*. [23] proposed a multiphysics model to demonstrate energy transfer from a spherical acoustic source to a piezoelectric disk in fixed-free boundary conditions. The linear model was validated with finite element (FE) simulations and later experimentally validated for free-free boundary conditions [21]. Ozeri *et al*. [6] also used a linear acoustic model to estimate the power transfer through acoustic waves propagating from planar transducers, and traveling in tissue at 673 kHz to actuate a piezoelectric disk. The assumptions of the linear acoustic-electroelastic model employed in most of the works in UPT fail to hold when acoustic nonlinearities, and/or piezoelectric geometric and material nonlinearities are triggered. Operating with focused sources at high acoustic intensities (such as in HIFU), high acoustic frequencies, or in mediums having a high coefficient of nonlinearity such as tissues, makes the propagating acoustic waves nonlinear [31, 36-39]. On the other hand, piezoelectric geometric and material nonlinearities become dominant under large strains. In such cases, it becomes necessary to account for acoustic and structural nonlinearities for accurate predictions of the receiver responses. This work focuses on acoustic nonlinearities by taking into account the effects of diffraction, absorption, and nonlinearity in the medium on the propagating waves. To model acoustic nonlinearities from a focusing source, Khokhlov-Zabolotskaya-Kuznetsov (KZK) equation is used [40]. The equation is based on parabolic approximation, and is applicable for directional sound beams and focused transducers with limited aperture angles [41, 42]. This equation models forward wave propagation and accounts for nonlinear wave distortion in a direction normal to the propagation plane.

Another limitation of the current UPT modeling efforts arises due to the finite aspect ratio (diameter-to-thickness ratio) of the receiver/transmitter disks. The advantage of considering finite-size disks is the wide range of use of these disks in different applications ranging from proof-of-concept lab experiments in ultrasonic transcutaneous energy transfer for powering implanted devices [6, 7] to piezoelectric acoustic-electric power transfer for metal walls [20, 43-45]. Studies have reported that the assumption of piston-like vibration mode where the disk vibrates only in the thickness direction, does not hold for disks with aspect ratios between 0.1-20 [43]. This is primarily due to the motion of the disks occurring both in radial and thickness directions, as compared to the conventional assumption of only thickness direction motion. Thus, the derivation of a closed-form solution for a finite-size piezoelectric disk acting as a transmitter or a receiver becomes very complicated. Various works have used different approaches based on elasticity theory, plate theory, and numerical methods to study the structural response of transmitter disks under electrical excitation [43, 46, 47]. However, these modeling efforts are scarce for disks acting as receivers in an acoustic field. For the case of finite-size receivers, acoustic-structure interaction effects arising from the reflected and scattered acoustic field also need to be accounted for by the existing models, since these effects influence the non-planar motion of the disk and vice-versa. Consequently, this issue is addressed in this work using an FE based approach to formulating the acoustic-piezoelectric structure interaction problem.

Based on the above-mentioned limitations/challenges in the current UPT modeling efforts, this study aims to demonstrate a novel concept involving the acoustic-structure interaction effects of the HIFU nonlinear acoustic field on the response of a finite-size piezoelectric receiver. It is assumed that the strains produced in the receiver due to acoustic excitation are small such that structural or geometrical nonlinearities in the disk are not triggered. Following this, in section 2, the KZK equation is used to estimate sound pressure on a piezoelectric disk. The structural response of the disk is formulated using an FE approach. Based on boundary conditions, the acoustic-structure interaction is quantified through a coupling matrix, which gives information about the reflection and scattering of the pressure field in the presence of the disk. The FE model is implemented through COMSOL Multiphysics®. Experimental results and model validation are presented in section 3. Characteristics of the nonlinear acoustic field and the response of a finite-size receiver are first investigated individually, and then the combined system is analyzed to understand the interaction of the two physics. A summary of the power output characteristics and conclusions is given in section 4.

## 2. Theory

### 2.1. Nonlinear acoustic-electroelastic theory

The coupled acoustic-electroelastic multiphysics of high-intensity focused UPT involves the estimation of the focused ultrasound (FU) field from a HIFU transducer, and wave interaction with the receiver to predict the mechanically induced vibrational and electrical responses. First, to understand the wave kinematics, the KZK equation is used to estimate the acoustic pressure field on a piezoelectric disk submerged in a fluid domain. The disk is placed in a way such that the excitation is along the polarity direction, i.e., thickness direction, as shown in Fig. 1. Although in this work, the KZK equation predicts the pressure field only in the fluid domain, it can be easily extended to study the acoustic pressure field in a multi-domain environment, as demonstrated in [32]. The multi-domain approach is particularly useful when the receiver disk is placed in a heterogeneous domain such as inside the human body, where acoustic waves pass through multiple layers of skin, tissue, and muscles.

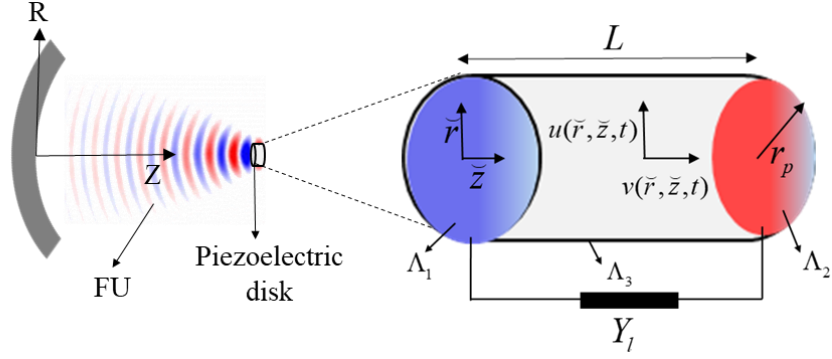

Fig. 1. Schematic representation of FU actuated piezoelectric receiver in a UPT system; the receiver is located in the focal area of the transducer in free-free boundary conditions.

The KZK equation incorporates the effects of absorption, diffraction, and nonlinearity of the medium on the wave propagation. It is expressed as [40, 48-50]

$$\frac{\partial^2 p}{\partial t \partial z} = \left(\frac{c}{2}\nabla_r^2 + \frac{\xi}{2c^3}\frac{\partial^3}{\partial t^3}\right)p + \frac{\beta}{2\rho c^3}\frac{\partial^2 p^2}{\partial t^2} \qquad (1)$$

where $p$ is the sound pressure at the observation point, $t$ is the time and $z$ is the wave propagation distance in the axial direction $Z$. The parameters $\rho$, $c$ and $\xi$ are the density, speed of sound, and diffusivity of the acoustic medium, respectively. The Laplacian operator $\nabla_r^2$ is defined as $\nabla_r^2 = \partial^2/\partial r^2 + (1/r)\partial/\partial r$, where $r$ is the radial distance in the radial direction $R$. The first, second, and third terms on the right-hand side of Eq. (1) represent diffraction, absorption, and nonlinearity effects of the medium on the wave, respectively. The parameter $\beta$ is the coefficient of the nonlinearity of the medium. For analyzing in the frequency domain, acoustic pressure is expressed as $p = (1/2)\sum_{n=1}^{\infty} u_n[r,z]e^{in\omega_0 t} + u_n^*[r,z]e^{-in\omega_0 t}$, where $\omega_0$ is the source angular frequency, $u_n$ is the complex pressure amplitude dependent upon spatial coordinates, $i$ is an imaginary number, and the operator $(\ )^*$ signifies the complex conjugate of the $n^{th}$ harmonic. Normalizing the quantities of pressure as $\bar{p}_n = u_n/p_0$, where $p_0$ is the characteristic source pressure, axial distance $z$ with focal distance $D$, as $\bar{z} = z/D$, and radial distance with source radius $a$, as $\bar{r} = r/a$, Eq. (1) in the frequency domain is rewritten as [51]

$$\frac{\partial \bar{p}_n}{\partial \bar{z}} = \left(-\frac{i}{4nG}\nabla_{\bar{r}}^2 - \bar{\mu}_n\right)\bar{p}_n + \frac{inN}{4}\sum_{m=1}^{\infty}\bar{p}_m\left(\bar{p}_{n-m} + 2\bar{p}_{m-n}^*\right) \quad n=1,2... \qquad (2)$$

The dimensionless parameter $G$ accounts for focusing gain, $G = \omega_0 a^2/2cD$, and $\bar{\mu}_n$ is a complex number denoting absorption. The nonlinearity parameter $N$ is the ratio of the focal length to the shock formation distance. It is defined as $N = D\beta\omega_0 p_0/\rho c^3$. The source condition is assumed to be a sinusoidal harmonic excitation of a uniform circular piston with a phase difference to account for the focusing effect. The corresponding boundary conditions for the KZK equation are [31]

$$\bar{p}_1[\bar{r},0] = \sin[\omega_0 \tau + G\bar{r}^2], \ \bar{r} \leq 1$$
$$= 0, \ \bar{r} > 1 \tag{3}$$

$$\bar{p}_n[\bar{r},0] = 0 \quad n = 2,3 \tag{4}$$

$$\bar{p}_n[\infty,\bar{z}] = 0 = \frac{\partial \bar{p}_n}{\partial \bar{r}}[0,\bar{z}] \quad n = 1,2... \tag{5}$$

Here, the variable $\tau$ is known as the retarded time and is given as $\tau = t - z/c$. In Eq. (2), absorption accounts for the thermoviscous attenuation leading to loss of energy in a propagating wave. Absorption is represented by $\bar{\mu}_n$, which is a complex number whose real part represents attenuation and the imaginary part gives dispersion relation. The attenuation coefficient, denoted by $\alpha$, follows an arbitrary power-law dependence with frequency, such that $\alpha[\omega] = \alpha_0 |\omega|^\nu$, where $\omega$ is the angular frequency of the propagating wave. The variables $\alpha_0$ and $\nu$ are constants and $0 \leq \nu \leq 2$ [52]. For fluids, $\nu = 2$ is usually used. To maintain causality, the dispersion relation follows the Kramers-Kronig relation [52].

Nonlinearity in the medium leads to the generation of higher harmonics making the wave distorted as it moves through the medium. The coefficient of nonlinearity is defined as $\beta = 1 + B/2A$ where $B/A$ is the nonlinearity parameter given as [40]

$$\frac{B}{A} = \frac{\rho_0}{c_0^2}\left(\frac{\partial^2 p}{\partial \rho^2}\right)\bigg|_{g=g_0} \tag{6}$$

where $g$ is the specific entropy of a state, $p = p[\rho, g]$, and $(\ )_0$ denotes a quantity in the unperturbed state.

An operator splitting method is used to numerically solve Eq. (2), which involves solving for each of the three terms in the right-hand side of the equation separately at each spatial integration step [53]. A hybrid time-frequency domain approach is adopted to solve the linear terms in the frequency domain using second-order finite difference methods. The nonlinear term is solved in the time domain using the upwind method [54]. For accurate modeling of the nonlinear pressure field, up to 128 harmonics are considered. A detailed explanation of the solving technique is given in Ref. [39].

The KZK equation-predicted acoustic pressure is used to estimate the force on a piezoelectric disk that is located in the focal region of the transducer. The direction of applied force is in the direction of wave propagation, as illustrated in Fig. 1. To model the electroelastic response, a disk of thickness $L$, density $\rho_p$, and radius $r_p$ connected to a circuit with load resistance $R_l$ ( $Y_l = 1/R_l$ for purely resistive load in Fig. 1 ) is considered in free-free boundary conditions. The variable $\breve{r}$ represents the radial distance from the center of the disk in the radial direction, $R$ and, $\breve{z}$ is the distance along the thickness of the disk in the axial direction, $Z$ (Fig. 1). The receiver is assumed to be transversely isotropic. Thus, it has an axisymmetric response. The displacements of the disk in radial and axial directions are denoted by $u[\breve{r},\breve{z},t]$ and $v[\breve{r},\breve{z},t]$, respectively. The coupled equations for the piezoelectric response are derived using Hamilton's principle, given as [55, 56]

$$\int_{t_1}^{t_2}(\delta T - \delta U + \delta W_{nc})dt = 0 \tag{7}$$

where the operator $\delta(\ )$ represents the variation of a quantity. The variables $T$, $U$, and $W_{nc}$ denote kinetic energy, potential energy, and work done by non-conservative or external forces, respectively. The potential energy of the receiver in the cylindrical coordinate system assuming zero free charges and zero initial potential [56], is

$$U = \int_{dV}\left(\frac{1}{2}\mathbf{S}^t Y \mathbf{S} - \mathbf{E}^t e \mathbf{S} - \frac{1}{2}\mathbf{E}^t \varepsilon \mathbf{E}\right)dV \tag{8}$$

where $Y$, $e$ and, $\varepsilon$ are the $6\times 6$ elastic modulus at the constant electric field, $3\times 6$ piezoelectric coupling, and $3\times 3$ permittivity matrices for isotropic piezoelectric materials [56], given in the appendix. The superscript $(\ )^t$ denotes the transpose of a quantity. The strain, $\mathbf{S}$ and, electric field, $\mathbf{E}$, vectors are defined as

$$\mathbf{S} = \begin{pmatrix} S_{\breve{r}} \\ S_{\varphi} \\ S_{\breve{z}} \\ 2S_{\breve{r}\varphi} \\ 2S_{\breve{r}z} \\ 2S_{\varphi z} \end{pmatrix} = \begin{pmatrix} \frac{\partial}{\partial \breve{r}} & 0 \\ \frac{1}{\breve{r}} & 0 \\ 0 & \frac{\partial}{\partial \breve{z}} \\ 0 & 0 \\ \frac{\partial}{\partial \breve{z}} & \frac{\partial}{\partial \breve{r}} \\ 0 & 0 \end{pmatrix}\begin{pmatrix} u \\ v \end{pmatrix} = [B]\mathbf{W} \text{ and } \mathbf{E} = \begin{pmatrix} E_{\breve{r}} \\ E_{\breve{z}} \end{pmatrix} = -\begin{pmatrix} \frac{\partial}{\partial \breve{r}} \\ \frac{\partial}{\partial \breve{z}} \end{pmatrix}\bar{V} = -[\mathrm{L}]\bar{V} \tag{9}$$

where $\bar{V}$ is electric potential, $\mathbf{W}$ is the displacement vector given as $\mathbf{W} = (u\ v)^t$ and, $[B]$ and $[L]$ are differential operator matrices. Taking variation, Eq. (8) becomes

$$\delta U = \int_{dV}\left(\delta\mathbf{S}^t Y \mathbf{S} - \delta\mathbf{E}^t e \mathbf{S} - \delta\mathbf{S}^t e \mathbf{E} - \delta\mathbf{E}^t \varepsilon \mathbf{E}\right)dV \tag{10}$$

The variation of the kinetic energy of the disk is

$$\delta T = \int_{dV}\delta\dot{\mathbf{W}}^t \rho_p \dot{\mathbf{W}}dV \tag{11}$$

where an overdot represents differentiation with respect to time. The variation of work done by the external force of acoustic pressure, damping, and electrical energy is

$$\delta W_{\text{nc}} = -\int_{\Lambda} \mathbf{p}_{\text{ext}} A_f \left.\delta \mathbf{W}^t\right|_{\Lambda} d\varsigma - \int_{dV} c_s \delta \mathbf{W}^t \dot{\mathbf{W}} dV - \int_{dV} Q \delta \overline{V} dV \qquad (12)$$

Here, the pressure exerted by sound, $\mathbf{p}_{\text{ext}}$, is integrated only over the acoustic-structure interaction surface, $\Lambda$, with a total surface area $A_f$. The variable $\Lambda$ is defined as $\Lambda = \Lambda_1 \cup \Lambda_2 \cup \Lambda_3$ (Fig. 1). The variable $c_s$ is the structural damping coefficient and $Q$ is the net electric charge in the disk. Substituting Eqs. (10-12) in Eq. (7) gives the equation of motion as

$$\int_{t_1}^{t_2} \left( \begin{array}{c} \int_{dV} \left( \delta \dot{\mathbf{W}}^t \rho_p \dot{\mathbf{W}} - \delta \mathbf{S}^t Y \mathbf{S} + \delta \mathbf{E}^t e \mathbf{S} + \delta \mathbf{S}^t e \mathbf{E} + \delta \mathbf{E}^t \varepsilon \mathbf{E} - \delta \mathbf{W}^t c_s \dot{\mathbf{W}} - Q \delta \overline{V} \right) dV \\ - \int_{\Lambda} \mathbf{p}_{\text{ext}} A_f \left.\delta \mathbf{W}^t\right|_{\Lambda} d\Lambda \end{array} \right) dt = 0 \qquad (13)$$

To ensure continuity at the boundary interface, the normal velocity of the structural boundary should be identical to the fluid velocity along the surface normal. Secondly, the acoustic force acting on the structure should be equal and opposite to the force exerted by the structure on the fluid. These two conditions are expressed as [57]

$$[B] p.\breve{\mathbf{N}} = -\rho \ddot{\mathbf{W}}.\breve{\mathbf{N}}$$
$$p.\breve{\mathbf{N}} = -\rho_p \ddot{\mathbf{W}}.\breve{\mathbf{N}} \qquad (14)$$

where $\breve{\mathbf{N}}$ is the unit normal to the surface of the boundary.

### 2.2. Finite element analysis

Previous works have developed closed-form solutions to estimate the electromechanical response of the disk under acoustic excitation [21, 23]. Such a closed-form approach is generally possible when the disk can be assumed to respond in a piston-like motion. In these scenarios, any shear effects and radial motion of the disk can be neglected. However, disks with finite ratios, such as discussed in Ref. [43], and this work, do not show a piston-like motion. Therefore, the general convention of assuming spatial dependence of displacement on a single cylindrical coordinate is not valid for such cases. The response of finite-size disks depends both on radial and thickness directions and is complicated to capture analytically [43]. Thus, numerical techniques are adopted to predict their responses accurately. The FE technique formulation will be briefly discussed here for piezoelectric structures [43, 58, 59]. Assuming the piezoelectric disk as an axisymmetric system of discrete non-overlapping elements (denoted by superscript $e$), the displacement and electric fields for an element are expressed as

$$\mathbf{W} = [\phi] \mathbf{N}^e \text{ and } \overline{V} = [\phi_e] \mathbf{R}^e \qquad (15)$$

Where $[\phi]$, and $[\phi_e]$ are $2 \times n$ and $1 \times n$ matrices of quadratic Lagrange shape functions respectively. The vectors $\mathbf{N}^e$ and $\mathbf{R}^e$ are nodal displacements and electric potential vectors respectively, such that $N_i^e = (u_i^e \ v_i^e)^t$ and $R_i^e = (\bar{V}_i^e)$ at the $i^{th}$ node of the element. Substituting Eq. (15) in Eq. (13) gives

$$\int_{t_1}^{t_2} (\delta \mathbf{N}^e)^t \left( M^e \ddot{\mathbf{N}}^e + K^e \mathbf{N}^e + k^e \mathbf{R}^e + d^e \dot{\mathbf{N}}^e - \mathbf{F}^e \right) + (\delta \mathbf{R}^e)^t \left( -(k^e)^t \mathbf{N}^e + \hat{\varepsilon}^e \mathbf{R}^e - \hat{\mathbf{Q}}^e \right) dt,$$

which yields the equation of motion as

$$M^e \ddot{\mathbf{N}}^e + K^e \mathbf{N}^e + k^e \mathbf{R}^e + d^e \dot{\mathbf{N}}^e - \mathbf{F}^e = 0 \qquad (16)$$

along with electrical circuit equation

$$-(k^e)^t \mathbf{N}^e + \hat{\varepsilon}^e \mathbf{R}^e = \hat{\mathbf{Q}}^e \qquad (17)$$

where $M^e = \int_{V_e} [\phi]^t \rho_p [\phi] dV$, $K^e = \int_{V_e} ([B][\phi])^t Y ([B][\phi]) dV$, $k^e = \int_{V_e} ([B][\phi])^t e ([L][\phi]) dV$,

$d^e = \int_{V_e} [\phi]^t c_s [\phi] dV$, $\hat{\varepsilon}^e = \int_{V_e} [L(\phi_e)]^t \varepsilon [L(\phi_e)] dV$ and $\hat{\mathbf{Q}}_e = [\phi_e]^t \mathbf{Q}_e$. Here, $\mathbf{Q}_e$ is the net electric charge vector at each node in the element. The external forcing is given as $\mathbf{F}^e = \int_{\Lambda_e} [\phi]^t A_f^e \mathbf{p}_{ext}^e$. The equations of all elements are assembled to obtain global equations with the help of connectivity matrices [59]. Assuming boundary conditions of axisymmetry, the coupled system of the equation of motion and electrical circuit equation is given as

$$\begin{pmatrix} M & 0 \\ 0 & 0 \end{pmatrix} \begin{Bmatrix} \ddot{\mathbf{N}} \\ \ddot{\mathbf{R}} \end{Bmatrix} + \begin{pmatrix} d & 0 \\ 0 & 0 \end{pmatrix} \begin{Bmatrix} \dot{\mathbf{N}} \\ \dot{\mathbf{R}} \end{Bmatrix} + \begin{pmatrix} K & k \\ -(k)^t & \hat{\varepsilon} \end{pmatrix} \begin{Bmatrix} \mathbf{N} \\ \mathbf{R} \end{Bmatrix} = \begin{Bmatrix} \mathbf{F} \\ \hat{\mathbf{Q}} \end{Bmatrix} \qquad (18)$$

For the receiver shown in Fig. 1, the electrical power is measured by connecting a load resistance across the two radial surfaces perpendicular to the axis ($Z$ direction) of the disk, $\Lambda_1$ and $\Lambda_2$. These two surfaces are assumed to be equipotential surfaces. Considering these surfaces as electrodes with a potential difference $V_0$ across them, a current equivalent to $V_0/R_l$ passes through the disk. Assuming $\Lambda_2$ as ground, the potential vector of all nodes on $\Lambda_1$, $\mathbf{V_0}$, contributes to the electric power across the load. The charge accumulated at the free surface, $\bar{\mathbf{Q}}$, is expressed as $\bar{\mathbf{Q}} = \int_{t_1}^{t_2} \mathbf{V_0}/R_l \ dt$. The equation of motion coupled with the electrical circuit equation now becomes

$$\begin{pmatrix} M & 0 \\ 0 & 0 \end{pmatrix} \begin{Bmatrix} \ddot{\mathbf{N}} \\ \ddot{\mathbf{R}} \end{Bmatrix} + \begin{pmatrix} d & 0 \\ 0 & 0 \end{pmatrix} \begin{Bmatrix} \dot{\mathbf{N}} \\ \dot{\mathbf{R}} \end{Bmatrix} + \begin{pmatrix} K & k \\ -(k)^t & \hat{\varepsilon} \end{pmatrix} \begin{Bmatrix} \mathbf{N} \\ \mathbf{V_0} \end{Bmatrix} = \begin{Bmatrix} \mathbf{F} \\ \bar{\mathbf{Q}} \end{Bmatrix} \qquad (19)$$

## 2.3. Acoustic-structure interaction

For a disk immersed in a fluid and excited by an acoustic source, the motion of the disk will be influenced by the acoustic medium loading on the disk. In parallel, the acoustic pressure field near the disk will also be influenced by the vibration of the disk surfaces. In such cases, the equations of motion are derived similarly using Hamilton's principle. However, an additional term accounting for the work done by acoustic pressure on the disk surface is added. In this work, since the output response of the disk is of interest, the acoustic-structure interaction problem is formulated only for the disk [57].

In Eq. (16), the forcing term $\mathbf{F}^e$ represents external forcing effects on an element of the disk. If fluid elements are also discretized such that $\mathbf{p}^e_{ext} = [g]\mathbf{p}^e_t$, where $[g]$ is a quadratic Lagrange shape function and $\mathbf{p}^e_t$ is a nodal pressure vector for a fluid element, the forcing is then expressed as $\mathbf{F}^e = \int_{\Lambda_e} [\phi]^t A^e_f [g] \mathbf{p}^e_t d\Lambda = [H] \mathbf{p}^e_t$. Here, $[H]$ is known as the acoustic-structure coupling matrix.

For the disk shown in Fig. 1, the total external force on a disk element can be further decomposed as the contributions from incident pressure from acoustic source, $[g]\mathbf{p}^e_i$, the reflected pressure when the disk acts like a rigid body (no motion of surface), $[g]\mathbf{p}^e_r$, and the scattered/radiated pressure due to vibration of disk elements, $[g]\mathbf{p}^e_{rad}$. The external force, $\mathbf{F}^e$, now becomes [57]

$$\mathbf{F}^e = [H]\left(\mathbf{p}^e_i + \mathbf{p}^e_r + \mathbf{p}^e_{rad}\right) \tag{20}$$

When the disk acts as a rigid body, the total pressure on the disk is the summation of only two components, reflected and incident pressure. This sum is known as block pressure $\mathbf{p}^e_{blo}$ and is equal to $\mathbf{p}^e_{blo} = \mathbf{p}^e_i + \mathbf{p}^e_r$ [57]. Consequently, the total external force on an element is re-written as $\mathbf{F}^e = [H]\left(\mathbf{p}^e_{blo} + \mathbf{p}^e_{rad}\right)$. Accounting for these pressure effects individually and assembling to get a global equation of motion coupled with the electrical response, Eq. (19) becomes

$$\begin{pmatrix} M & 0 \\ 0 & 0 \end{pmatrix}\begin{Bmatrix} \ddot{\mathbf{N}} \\ \ddot{\mathbf{R}} \end{Bmatrix} + \begin{pmatrix} d & 0 \\ 0 & 0 \end{pmatrix}\begin{Bmatrix} \dot{\mathbf{N}} \\ \dot{\mathbf{R}} \end{Bmatrix} + \begin{pmatrix} K & k \\ -(k)^t & \hat{\varepsilon} \end{pmatrix}\begin{Bmatrix} \mathbf{N} \\ \mathbf{V_0} \end{Bmatrix} - \begin{pmatrix} H & 0 \\ 0 & 0 \end{pmatrix}\begin{Bmatrix} \mathbf{P}_{rad} \\ 0 \end{Bmatrix} = \begin{Bmatrix} \mathbf{F}_{blo} \\ \overline{\mathbf{Q}} \end{Bmatrix} \tag{21}$$

where $\mathbf{F}_{blo} = [H]\mathbf{p}_{blo}$. To understand the effects of the radiated pressure field on the vibrational response of the disk, the radiation pressure can be further expressed in terms of radiation impedance matrix, $z^e_{wf}$, and associated velocity of the disk such that $\int_{\Lambda_e} [g]\mathbf{p}^e_{rad} d\Lambda = -\int_{\Lambda_e} [z^e_{wf}][\phi]\dot{\mathbf{N}}^e d\Lambda$ [60]. Here, the radiation impedance is composed of two components; a resistive part, $R^e_{wf}$, and an imaginary part, $X^e_{wf}$. The resistive component contributes to the damping effect, whereas the imaginary portion adds to the inertia and shifts the natural frequency, $\omega_r$, of the disk. Moreover, for a harmonic response of the disk, the reactive term of the fluid wave impedance can further be expressed as an added mass, given by $[X^e_{wf}]\dot{\mathbf{N}}^e = \left[\frac{X^e_{wf}}{\omega_r}\right]\ddot{\mathbf{N}}^e = [M^e_{wf}]\ddot{\mathbf{N}}^e$. When these two radiation impedance components are assembled and incorporated into the global equation of motion, the first row of Eq. (21) becomes

$$[M + M_{wf}]\ddot{\mathbf{N}} + [d + R_{wf}]\dot{\mathbf{N}} + K\mathbf{N} + k\mathbf{V_0} = \mathbf{F_{blo}} \qquad (22)$$

Eq. (21) is solved using COMSOL Multiphysics®, with the setup comprising of a piezoelectric disk in a finite water domain, placed at the focal point of the transducer. The water domain is of radius 70 mm and lined with a perfectly matching layer of 3 mm to simulate an infinite medium. The disk is poled such that the radial surface enclosed by $\Lambda_2$ serves as the ground, while the opposite surface gives the net potential. This ungrounded surface faces the incoming acoustic waves. The disk is also attached to a 1-ohm resistance to estimate the output power at short-circuit natural frequencies corresponding to different modes. A mesh of quadratic Lagrangian elements is chosen with the maximum element size limited to six elements per acoustic wavelength. For time-domain simulations, a generalized alpha solver with a manual time-step size corresponding to the highest frequency component in the acoustic field is used.

## 3. Experimental validation and acoustic-structure interaction characterization

Experiments are conducted with an H-104-4A SONIC Concepts HIFU transducer mounted on one side of a 61.5 × 31.8 × 32.5 cm³ water tank, as shown in Fig. 2(a). The water tank is filled with deionized water to avoid any electrical short-circuiting. A Precision Acoustics 1 mm needle hydrophone measures the acoustic field of the transducer using a TBS2000 Series Tektronix oscilloscope, Fig. 2(a). The readings from the oscilloscope are recorded using a built-in MATLAB interface. The hydrophone is connected to the oscilloscope via a DC coupler, which conditions the hydrophone signal and also acts as a power supply. The hydrophone is mounted on a positioning system, which scans the acoustic field in axial and radial directions with respect to the transducer, to acquire pressure measurements. The water tank is lined with Aptflex F28 absorber sheets (purchased from Precision Acoustics Ltd.) on the bottom of the tank and the two side walls of the transducer, to prevent boundary wall reflections. The HIFU transducer is operated at 0.5 MHz with a 100 µs burst signal and 10 ms of burst period for different input electrical power to the HIFU amplifier.

### 3.1. Acoustic parameter identification and model validation

To model the acoustic field from the HIFU source used in experiments, the knowledge of the effective radius of curvature, and the operational aperture of the transducer is needed. The values of these parameters stated by the manufacturer do not allow the transducer to be modeled as a uniform piston source. Transducer housing, surface waves, and inhomogeneity in the piezoelectric elements of the transducer can distort the source vibration [61]. Therefore, effective values of source curvature and aperture are determined first.

The experimental setup as shown in Fig. 2, is used to measure pressure from the transducer at a low input power. The measurements are taken along the axial axis, and in the radial plane perpendicular to the axial axis at the focal point. At low input power, acoustic nonlinearities are negligible. Therefore, direct relationships to estimate the effective source geometry can be used. Measurements are taken and compared with the linear model ($\beta \approx 0$) in Eq. (1). Values of source aperture and radius of curvature are varied in the model to obtain the best agreement with the experimentally obtained axial and radial pressure fields. It is determined that the effective value of the radius of curvature is $D = 8$ cm, and the source aperture is $2r = 7.4$ cm. Based on these values, the model in Eq. (1) is validated with experimental observations and FE simulations in the water domain. The value of speed of sound, $c$, and water density, $\rho$, are considered as 1483 m/s and 1000 kg/m³ in the model. The FE simulation setup remains the same as described in the COMSOL implementation in section 2, but without the piezoelectric disk. Fig. 3 shows a good agreement between the KZK-calculated, FE, and experimental values of acoustic pressure in axial and transverse directions, under linear propagation. In this figure, the origin $z = 0$ denotes the focal point on the axial axis, $Z$. The prefocal fluctuations in the pressure field as seen in the figure, arise from the interference pattern due to diffraction of acoustic waves emitted by a focused transducer [31, 41, 50, 61]. The slight discrepancy in these fluctuations between measured and calculated values is due to parabolic approximation in the KZK model [61]. This validation shows that a single-element HIFU transducer can be modeled as a uniformly vibrating source.

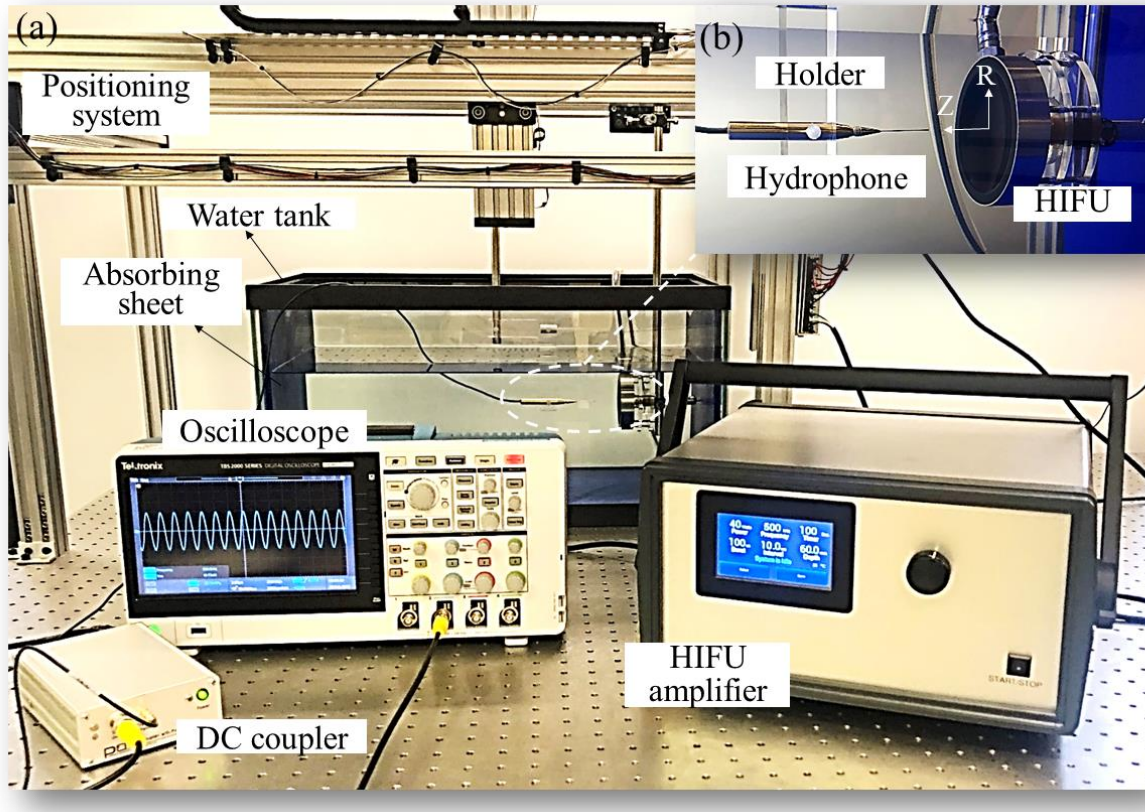

Fig. 2. (a) Experimental setup to measure the pressure field of a HIFU transducer in a water tank using a needle hydrophone. (b) A magnified image of the needle hydrophone inside the tank. The tip of the hydrophone traverses in axial and radial directions of the transducer to map the acoustic pressure field.

Having identified the effective characteristic values of the transducer geometry, it is important to find the pressure at the surface of the transducer. This involves estimating the conversion coefficient between the input voltage to the transducer and source pressure amplitude. Since it is difficult to measure the pressure at the source experimentally, an alternative analytical method is used [61]. Using the effective values of transducer geometrical parameters found above, linear focusing gain, $G$ in Eq. (2) is determined. The source pressure amplitude is then estimated as $p_0 = p_f/G$ where $p_f$ is the pressure at the focal point. For a voltage input, $V_{in} = 4$ V, to the HIFU amplifier, the source pressure value is calculated as $p_0 = 8$ kPa. The conversion coefficient, $\eta_{av}$, is then estimated to be $\eta_{av} = p_0/V_{in}$, with a value of approximately 2 kPa/V. The source pressure is found to linearly increase with input voltage and is estimated using $\eta_{av}$ for future measurements.

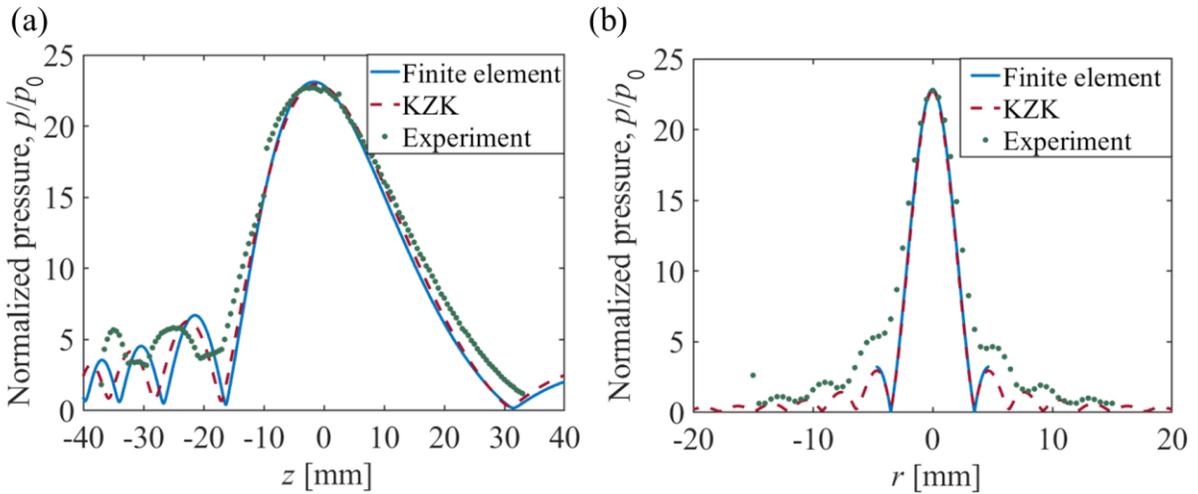

Fig. 3. Pressure field of a HIFU source on (a) axial and (b) radial axis ($V_{in} = 4$ V and $p_0 = 8$ kPa) in water at 0.5 MHz.

After calibrating the model to account for effective geometrical parameters and boundary conditions using linearized Eq. (1), the nonlinear acoustic model developed in section 2 is employed to predict the FU pressure field of the transducer. Figs. 4(a) and 4(b) show the experimental validation of the focal point pressure waveforms predicted by the KZK model with good agreement, at different source pressure, $p_0$, in time and frequency domain, respectively. A discrepancy can be observed between the experimental and KZK calculated values at higher frequency components for low source pressures in Fig. 4(b). This is because the amplitude of pressure for higher harmonics, obtained from experiments, is very low for low input power and close to noise. Besides, the finite size of the receiver compared to the excitation wavelength and its non-uniform sensitivity across frequencies also contributes to this discrepancy [61].

### 3.2. *Effects of acoustic parameters on acoustic nonlinearity*

The use of FU for different applications can lead to a strong interplay between diffraction, absorption, and nonlinearity effects on the wave under various conditions, such as a change in medium or source parameters. Such an interplay ultimately affects the characteristics of the propagating acoustic waves including amplitude, and the number of harmonics in the wave (section 2). Our previous work [39] conducted an in-depth analysis to study the influence of these effects on the sound pressure field at the focal point. A summary of our earlier work [39] is explained here, which will enhance the understanding of the interaction of the nonlinear acoustic field with piezoelectric structures discussed in the later sections, which is the focus of this work.

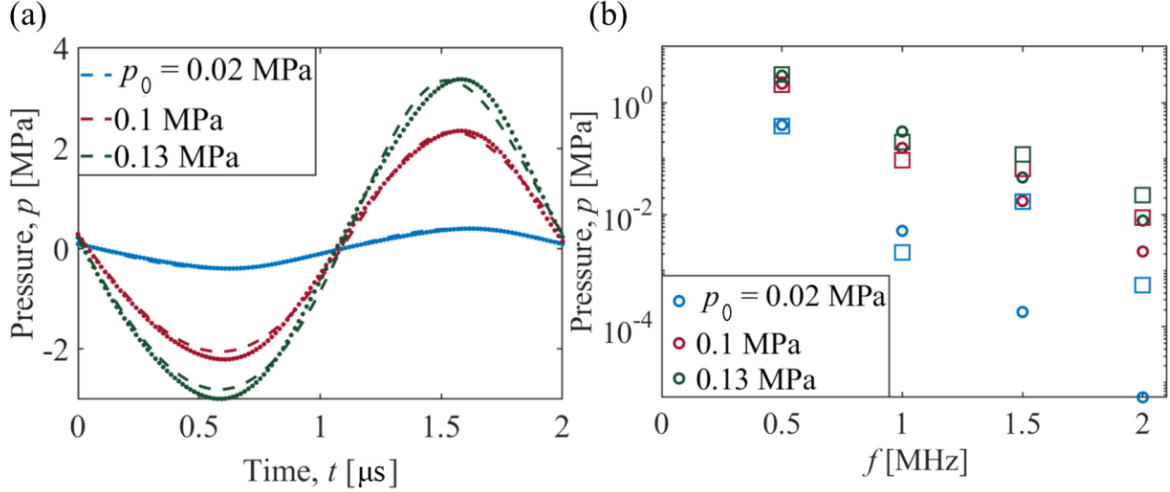

Fig. 4. Pressure waveforms at the focal point of the HIFU transducer operating at 0.5 MHz in (a) time and (b) frequency domain for various excitation levels in the water. Dashed and dotted lines in (a) represent KZK model calculations and experimental observations, respectively. Circular and square symbols represent KZK model calculations and experimental observations in (b), respectively.

The parameters that account for focusing gain, absorption and nonlinearity in the model are $G$, $\bar{A}$, and $N$, respectively, Eq. (2). Since the geometry of the transducer and its operating frequency are fixed for the experimental setup in this study, $G$ remains constant. The nonlinear effects in the medium, accounted by $N$, cause the propagation speed of the wave to vary from point to point resulting in the waveform peaks to travel faster as compared to troughs. This leads to distortion of the waveform and generation of higher harmonics. Due to the contribution of higher harmonics, the overall pressure amplitude at the focal point increases. From section 2, it is observed that $p_0$ and $\beta$ are two parameters that can change the value of $N$, Eq. (2). While $p_0$ depends on the transducer's input driving voltage (section 3.1), $\beta$ is a medium property and changes with different mediums. Figs. 5(a) and 5(b) show the variation of focal pressure waveforms obtained from the KZK model with changes in $p_0$ and $\beta$, respectively. An increase in the magnitude of both the parameters leads to a nonlinear increase in focal pressure amplitude; however, the mechanism in which they affect this pressure is different. While with an increase in $p_0$, the strength of each of the higher harmonics increases, Fig. 5(a), with amplification in $\beta$, the energy from the fundamental harmonic of the acoustic wave is transferred to higher harmonics, Fig. 5(b) [39], which grow under focusing effects.

As opposed to nonlinear effects, absorption effects in the medium cause loss of overall energy in the wave as it propagates. The absorption effects in Eq. (2) are accounted through a non-dimensional parameter, $\bar{A}_n$, denoting absorption of $n^{th}$ harmonic of the FU pressure field, defined as $\text{Re}[\mu_n] = \bar{A}_n = \bar{A}|\omega|^\nu$ where $\bar{A} = \alpha_0 D$. This expression states that attenuation is higher for higher harmonics. Thus, it counters the nonlinear effects since nonlinearity leads to the generation of higher harmonic components. Fig. 5(c) shows the change in the focal pressure field until $\bar{A}$ becomes significantly high (in the case of solids [62, 63]), and the pressure wave reduces to a low amplitude linear wave. This understanding of the variation in focal pressure gives an insight into the acoustic-structure interaction and the consequent electroelelastic response of the disk when the source strength or propagating medium changes.

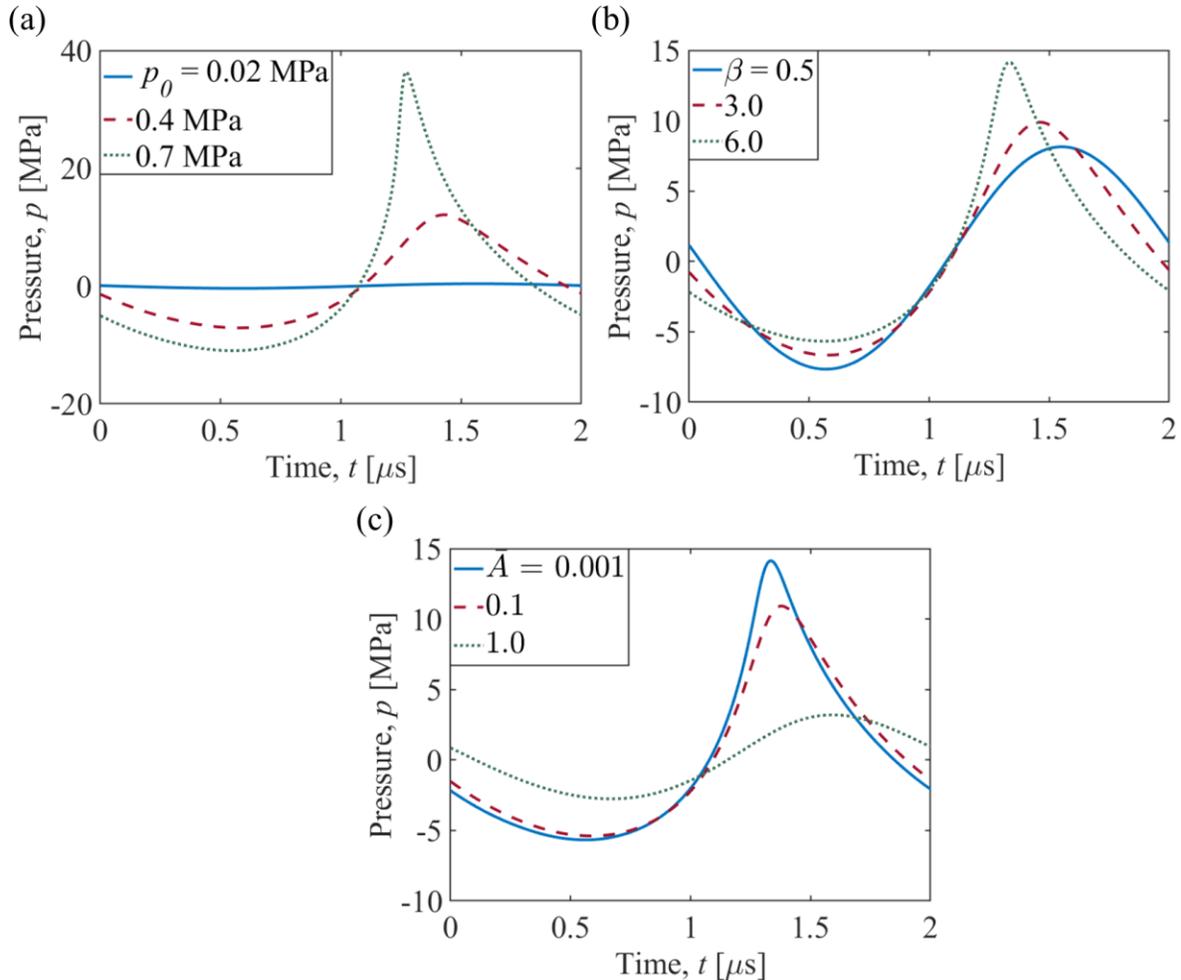

Fig. 5. Pressure waveform at the focal point in water domain, obtained from KZK model, for (a) various source pressure, $p_0$, at $\beta = 3.5$ and $\bar{A} = 5 \times 10^{-4}$, (b) various coefficient of nonlinearity, $\beta$, at $p_0 = 0.34$ MPa and $\bar{A} = 5 \times 10^{-4}$, and (c) various attenuation parameters, $\bar{A}$, at $p_0 = 0.34$ MPa and $\beta = 6$.

### 3.3. Experimental validation for a HIFU-UPT system

A UPT system consists of a transmitter generating acoustic waves incident on a receiver, both of which are immersed in the same or different mediums. In this work, the transmitter is the HIFU transducer and the receiver is a piezoelectric ceramic disk, APC760. The disk is made up of PZT-5A material manufactured by APC International, Ltd. In order to design a UPT system, impedance measurements of the disk are collected first to identify the disk's electroelastic parameters. The electrical impedance measurement of a freely hanging APC760 disk, suspended in the air with wires, is performed using an HP4192A impedance analyzer in the frequency range of 5 Hz - 3 MHz. Since the soldering of the wires occupies a very small region on the disk, the disk can be assumed to be in free-free boundary conditions and stress free. This assumption is consistent with previous observations [21, 45]. The FE formulation for electrical impedance in the air is then curve-fitted to the experimentally obtained impedance, by tuning the material properties of the disk given by the manufacturer. The impedance measurements in the air are used as a reference to find the disk properties because the fluid loading effects in air are negligible. The FE formulation for the electrical impedance of the disk in air can be derived from Eq. (21), where $\mathbf{p}_{rad}$ is negligible. Since the disk is acting as an actuator in impedance measurements, $\mathbf{F}_{blo} = 0$ and $\mathbf{V_0} = V^{in}e^{i\omega t}$, where $V^{in}$ is the driving voltage amplitude. The total charge, in this case, is $\bar{\mathbf{Q}} = \int_{t_1}^{t_2} I^{in} e^{i\omega t + \theta} dt$, where $I^{in}$ is the current passing through the disk. Thus, the FE formulation for the electroeleastic response of an actuating piezoelectric disk in the air is

$$\begin{pmatrix} M & 0 \\ 0 & 0 \end{pmatrix} \begin{Bmatrix} \ddot{\mathbf{N}} \\ \ddot{\mathbf{R}} \end{Bmatrix} + \begin{pmatrix} d & 0 \\ 0 & 0 \end{pmatrix} \begin{Bmatrix} \dot{\mathbf{N}} \\ \dot{\mathbf{R}} \end{Bmatrix} + \begin{pmatrix} K & k \\ -(k)^t & \hat{\varepsilon} \end{pmatrix} \begin{Bmatrix} \mathbf{N} \\ \mathbf{V_0} \end{Bmatrix} = \begin{Bmatrix} 0 \\ \bar{\mathbf{Q}} \end{Bmatrix} \quad (23)$$

Eq. (23) is implemented in COMSOL, with a disk surrounded by air in a 2-D axisymmetric model. The maximum mesh size is limited to six elements per wavelength. Eq. (23) is then used to identify electroelastic parameters using curve-fitting of the FE calculated impedance curve to the experimentally obtained impedance curve. The curve-fitting is performed for the short circuit frequency around 0.5 MHz since it is the resonant frequency of the HIFU transducer. Figs. 6(a) and S3(a) show the FE predicted impedance curves obtained after incorporating the identified parameters and the experimentally measured impedance in air. A slight discrepancy is observed at the open circuit frequency in Fig. 6(a), which can be due to the estimation of material properties based only on one frequency (0.5 MHz) or non-unique combination of material parameters used for curve-fitting. The identified parameters of the disk are reported in Table 1, where $\varepsilon_r$ is the permittivity of vacuum and $\zeta$ is the mass proportional Rayleigh damping. The mass proportional damping is related to Eq. (23) as $d = \zeta M$.

Using the properties mentioned in Table 1, the electrical impedance for APC760 disk is calculated in water. The FE formulation in Eq. (23) is modified to include the radiated pressure, $\mathbf{p}_{rad}$. The experimental setup to obtain the electrical impedance of the disk in water remains the same as the setup for air, however, the disk is now fully submerged in the water domain. Figs. 6(b) and S3(b) show a reasonable agreement between the impedance predicted from the FE with that of the experimental values. A vertical shift between the two curves is observed in Fig. 6(b), which can be due to a small change, i.e. $\pm 20\%$, of the capacitance of the APC760 disk after immersing in water. On comparing the impedance curves between air and water in Fig. 6, it is observed that the impedance amplitude decreases, and the short circuit resonant frequency shifts to the left when the disk is underwater. These behaviors can be attributed to the added damping and added mass effects due to fluid loading, respectively, as given in Eq. (22) [56]. It is also seen that the impedance calculated from the FE does not match the experimental values at other modes (Fig. S3), for both air and water. A primary reason is that the matching of the impedance curves obtained from two methods is performed only for one mode (around 0.5 MHz). Therefore, the material properties such as damping and electromechanical coupling, which are different for each mode, are not accounted for by the FE formulation and lead to the discrepancy with experimental values.

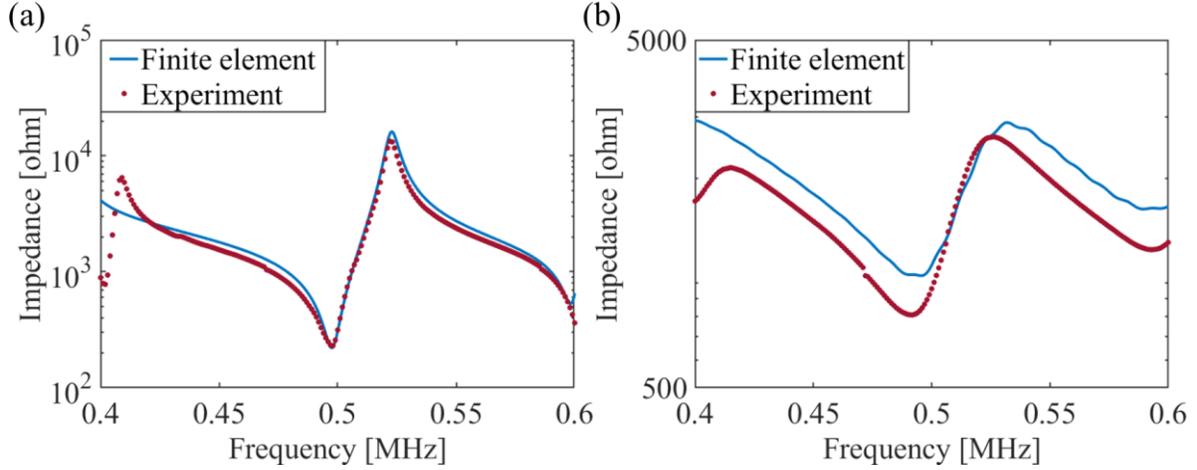

Fig. 6. Electrical impedance curves obtained from FE simulations (solid line) and experiments (dotted line) in (a) air and (b) water, for the APC760 disk.

Table. 1. Electroelastic properties of APC760 piezoelectric disk used in this study

| Property | Value |
|---|---|
| $\rho_p$ | 7700 kg/m$^3$ |
| $r_p$ | $4.75 \times 10^{-3}$ m |
| $L$ | $3.9 \times 10^{-3}$ m |
| $Y_{11} = Y_{22}$ | 148 GPa |
| $Y_{12} = Y_{21}$ | 105 GPa |
| $Y_{13} = Y_{31}$ | 106.82 GPa |
| $Y_{33}$ | 138.99 GPa |
| $e_{31} = e_{32}$ | $-1.5$ [C/m$^2$] |
| $e_{33}$ | 22.5 [C/m$^2$] |
| $e_{24} = e_{15}$ | 11.64 [C/m$^2$] |
| $\varepsilon_{11}/\varepsilon_r = \varepsilon_{22}/\varepsilon_r$ | 1130 |
| $\varepsilon_{33}/\varepsilon_r$ | 800 |
| $\zeta$ | 34000 [1/s] |
| $c_p$ | 4350 [m/s] |

With the developed understanding of the fluid loading effects on the disk in air and underwater, and identifying the material properties of the disk as found from experiments in air, the UPT system is designed. Fig. 7(a) shows the experimental setup, which is similar to the setup described in Fig. 2. However, in Fig. 7(a), the hydrophone is replaced by the APC760 disk. The disk is soldered using wires and mounted on the positioning system in the free-free boundary conditions. It is required that the focal point falls on the disk's leading surfaces since the acoustic field is concentrated at the focal point. To achieve this, two laser pointers are used as placeholders for the focal point of the HIFU, which is first located using the hydrophone. Once the hydrophone is removed, the lasers are used to pinpoint the focal point and to place the disk in the desired location; the green laser light can be seen in Fig. 6(a). The HIFU transducer sends pressure pulses with 10 ms of burst period having 100 µs of burst (50 cycles) at 0.5 MHz, to avoid any reflections and standing waves between HIFU and the disk. The voltage output of the acoustically excited disk is recorded using the MATLAB interface of the oscilloscope. A lock-in amplifier can also be used to analyze the disk response in the frequency domain. These experimental observations are used to validate the acoustic-structure interaction formulation developed in section 2, for APC760 disk. The FE model has already been validated for acoustic wave propagation in water with experiments and the KZK model, section 3.1, as well as for the response of disk for electrical actuation, Fig. 6. Consequently, the acoustic-structure FE model is implemented in COMSOL, according to the specifications described in section 2. Fig. 7(b) shows typical time histories of the voltage output across an electrical 1-ohm resistive load connected to the receiver disk, obtained from experimental measurements and the FE model. The acoustic excitation is maintained at a low-pressure level, $p_0 = 8$ kPa, such that acoustic or geometrical nonlinearities are not triggered. These results show that the proposed FE model is successfully able to capture the acoustic-piezoelectric interaction in the UPT system.

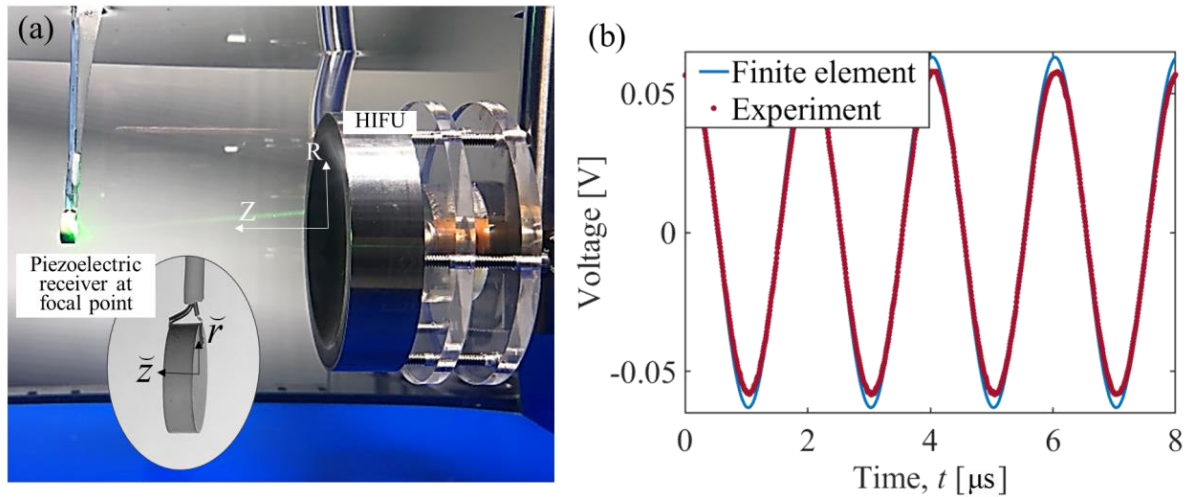

Fig. 7. (a) Experimental set up of the UPT system using HIFU source to excite a piezoelectric receiver at focal point (≈ 52 mm). The magnified image of the disk exposed to HIFU. The laser pointers fall on the top surface of the disk to mark the focal point of FU. (b) The voltage response of the receiver to signals of 10 ms of burst period having 100 µs of burst (50 cycles) at 0.5 MHz.

There are multiple reasons for adopting the FE approach in this work instead of using the reduced-order one-dimensional models [21, 23, 55], to study the structural response of the disk. A primary reason is the finite aspect ratio of the disk. As discussed in previous sections, disks with aspect ratios that are not very large or small (> 20 or < 0.1) [43], do not show the piston-like motion which these models assume. The mode shapes of such disks are the functions of both radial and axial directions [43, 64, 65]. Another reason for adopting an FE approach is that the structural resonant overtones are not harmonics for such disks. As seen from the eigenvalue analysis of disks having piston-type motion, the maximum voltage output occurs at the fundamental and its harmonic structural resonant frequencies [66]. However, due to the absence of pure-thickness modes in disks with finite aspect ratio, such an assumption does not hold. This can also be seen from the impedance curves in Fig. 8(a) and Fig. S3 and, power output response in Fig. S1. Two important observations can be made from these figures: (1) The power output is maximum from the receiver at those frequencies which coincide with the disk's short-circuit resonant frequencies in impedance measurements; (2) The frequencies at which maximum power is generated are not the harmonics of the fundamental frequency.

In addition, one another reason for adopting an FE approach is shown in Figs. 8(b1)-(b4). Figs. 8(b1)-(b4) show the normalized displacement profiles of the disk at its first four resonant frequencies, which have considerable power output upon acoustic excitation in water (Fig. S1). Mode shapes at all resonant frequencies are not shown due to brevity of space. A key observation to note in these figures is that the displacement profiles of the top and bottom surfaces of the disk at each frequency are not the same and differ in pattern from each other. This difference in patterns further complicates the attempts of developing reduced-order models or assuming a single universal displacement profile to describe the complete structural response. A possible reason for the difference between the displacement profiles of top and bottom surface displacements can be the unsymmetrical and nonuniform pressure and velocity profiles on the two leading surfaces. It is seen that the acoustic pressure field at both of these surface boundaries differs by a significant amount (Fig. S2). Since these pressure fields contribute to the radiation impedance, which in turn affects the motion of the disk, Eq. (22), it is possible that they affect the overall displacement pattern differently, for the top and bottom surfaces. Due to these various reasons, developing a comprehensive reduced-order model that accounts for all these effects becomes cumbersome. Thus, an FE based approach is needed for investigating the acoustic-electroelastic UPT system using a finite aspect ratio receiver.

### 3.4. *Electroelastic response of the piezoelectric receiver in the nonlinear acoustic field*

Having validated the FE formulation and understanding the structural response of the receiver disk for low acoustic excitation in a linear framework, the acoustic-structure interaction formulation, Eq. (21), is used to understand the effects of the nonlinear acoustic field on the piezoelectric disk responses. This investigation aims to understand the effects of multiple frequency components in the nonlinear acoustic wave on the receiver's electrical response. Such an investigation is particularly useful when these multiple frequency components become significant due to the inherent nature of the medium such as a high value of $\beta$ or due to high input power to the transducer, Fig. 5. A nonlinear acoustic wave can also be intentionally generated by sending a multi-frequency input signal to a transmitter. To model the nonlinear acoustic field, the experimentally-validated KZK model, in section 2, is used for the setup shown in Fig. 2. An advantage of using the KZK model as compared to solving the acoustic wave equation in the FE based time-domain simulation is the computational efficiency of the former method. The KZK model solves the acoustic wave equation using finite difference techniques, with accurate predictions, Fig. 4. The model predicted nonlinear pressure field at the focal point is then implemented as an external incident pressure input to the FE formulation, $\mathbf{p}_i = \sum \mathbf{p}_i^e$, Eq. (20), in the form of its Fourier components. This hybrid KZK-FE based model is then solved in COMSOL to estimate the electroelastic response of the disk in the time domain. Such an approach circumvents the heavy memory and computational time costs associated with solving the nonlinear wave equation in time domain using the FE approach.

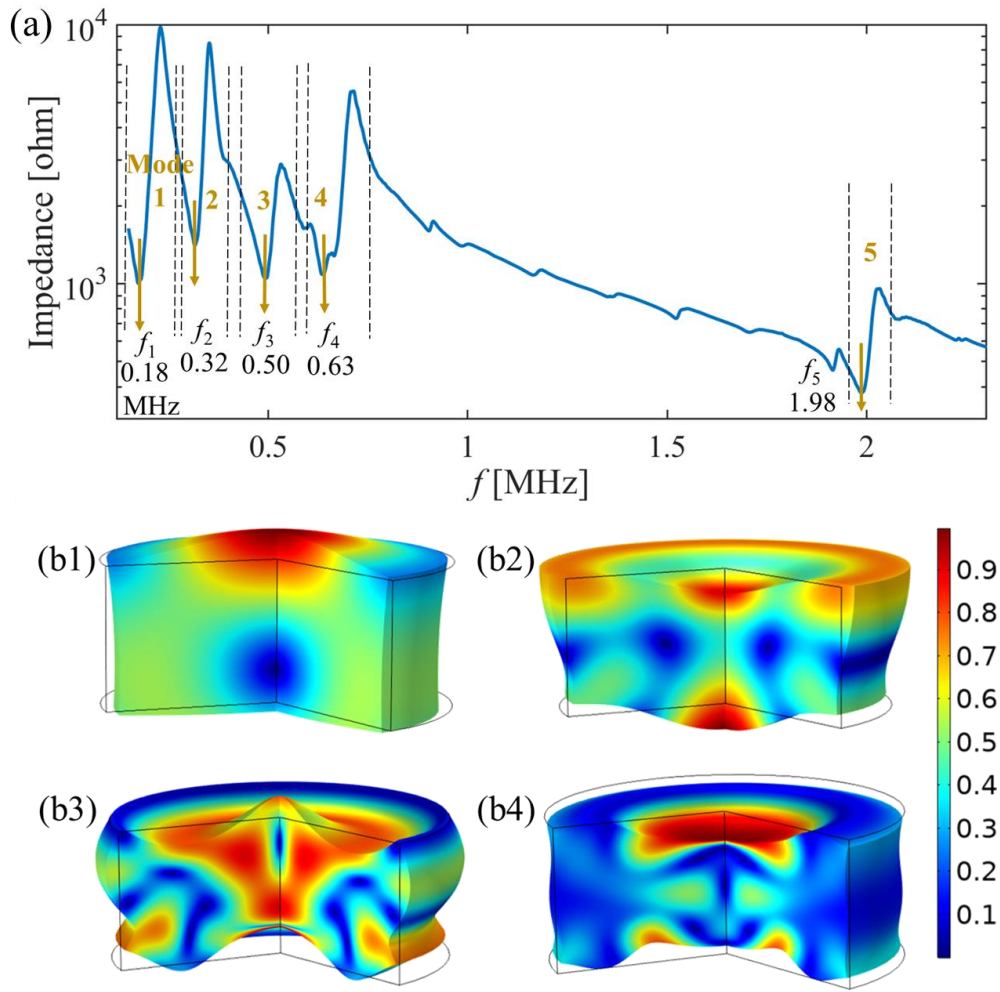

Fig. 8. (a) The electrical impedance curve of the APC 760 disk in water calculated from FE formulation include five dominant resonant modes of the APC760 piezoelectric disk. (b) Normalized displacement profiles of the acoustically excited disk obtained from the FE model for the first four short circuit resonant frequencies; (b1) 0.18, (b2) 0.32, (b3) 0.50, and (b4) 0.63 MHz. The normalization is done with respect to the maximum displacement in each case.

To assess the effects of the nonlinear acoustic field, four case studies are constructed. It is assumed in all four case studies that the geometrical and material nonlinearities of the piezoelectric disk are not triggered. Besides, a constant damping ratio is assumed in all the case studies since the time-domain simulations do not allow for real-time update of structural damping when the higher frequency components of the disk are excited. However, the damping ratio might be different for different modes and can be considered in the proposed model. The KZK-FE hybrid model is implemented in COMSOL for all case studies, with the simulation setup having an axisymmetric water domain of 70 mm in radius. A perfectly matched boundary layer of 3 mm is added. The disk is located at the center of the domain, and the KZK predicted focal pressure field is incident on the radial surface enclosed by $\Lambda_1$ boundary of the disk. The first case study, referred to as M1, is a benchmark study where a linear pressure field is incident on the disk; since in most of the works in the literature, acoustic nonlinearities are not accounted and a pressure wave solution based on linear Helmholtz equation is used [60]. Accordingly, the linear acoustic model, $\beta = 0$, in Eq. (2), is used and higher acoustic harmonic components are not considered. Figs. 9(a) and 9(b) show the KZK-FE model-based response of the disk for M1. The values of the non-dimensional parameters that are set as $N = 0$ and $\bar{A} = 0.001$, in Eq. (2). Due to the linear relation between acoustic and electroelastic model formulations, a mechanical excitation from linear acoustic wave gives a linear electrical response. Fig. 9(a) shows the KZK predicted incoming linear pressure wave, $p_i$, impinging on the disk and the resulting voltage response from the APC760 disk in the time domain. Fig. 9(b) shows the waveforms in corresponding frequency domain.

To understand the effect of higher harmonic components on the electroelastic response, the second case study, referred to as M2, is investigated. The acoustic source/input parameters remain the same as case study 1, however, the value of $\beta = 10$ is used to observe the nonlinear effects at a reasonable scale. Such high values of $\beta$ is usually seen in fat tissues [67]. As shown in Fig. 5(b), a high value of $\beta$ increases the focal pressure amplitude that grows nonlinearly with the generation of higher harmonics in the acoustic field. The value of non-dimensional parameter accounting for the attenuation also remains equal to the value used in M1. Figs. 9(c) and 9(d) show the KZK predicted incident nonlinear pressure wave on the disk and the corresponding voltage response. Only the first four Fourier components of the KZK predicted wave that have a considerable pressure amplitude are used as an input to the FE formulation. The reason for considering only the first four components is to save the computational cost. Since the structural response of the disk still needs to be solved in the time domain, it limits the maximum mesh size to one-sixth of the highest harmonic wavelength of the excitation force. Therefore, to account for such a fine mesh, the number of frequency components to evaluate this case study is restricted. It is seen that due to the presence of multi-frequency components, the voltage response also becomes nonlinear, obviously, having harmonics at the same frequencies as the incident acoustic field. However, an interesting feature in the response of the disk, Fig 9(d), is that the harmonic components of the voltage waveform are not monotonically decreasing. Because of a linear relation between excitation force and voltage response, this observation is contrary to the monotonically decreasing pattern of the acoustic pressure waveform, Fig. 9(c), although material nonlinearity is not pronounced. The reason for this non-monotonous variation can be inferred from the impedance curve of the disk (Fig. 8(a) and Fig. S3). The acoustic harmonic components which coincide with the structural resonant short circuit frequencies of the disk can elicit a significant voltage response, which in the M2 case study are the fundamental (~0.5 MHz) and the fourth harmonic (2 MHz) frequencies. It is seen in Fig. 9(d) that the second harmonic of voltage response is an order-of-magnitude lower, while the third harmonic is two orders lower than the fundamental component. Moreover, although the fourth acoustic harmonic has the lowest pressure amplitude, its corresponding voltage response is comparatively higher when compared to the voltage output from the second and third harmonics of acoustic excitation. Due to the additional presence of the fourth harmonic in voltage response, the average power output, $P_w$, 29.5 mW, is higher by 6.1% in this case study as compared to M1, 27.8 mW. However, the increase in the average power is not significant since only the fourth acoustic harmonic which has the smallest pressure amplitude as compared to other harmonics is contributing effectively to the receiver excitation. The average power is calculated as $P_w = V_{rms}^2 / R_1$, where $V_{rms}$ is the root mean squared (rms) voltage obtained from the time histories of voltage response.

From the observations of M2, it can be inferred that if a nonlinear acoustic field consists of frequency components coinciding with the resonant frequencies of the disk, the voltage response will increase. To verify this inference, the third case study, referred to as M3, is conducted. For M3, the same incident pressure field as in M2 is taken such that the total input energy from the HIFU transmitter remains the same; however, the frequencies of the higher harmonics are adjusted to coincide with the first four short-circuit resonant frequencies of the disk. Figs. 9(e) and 9(f) show this modified nonlinear pressure field and its corresponding voltage response. The average power, 51.5 mW, is significantly higher, by 75%, compared to the case studies M1 and M2. The relative magnitudes of the first to fourth frequency components of the voltage response in Fig. 9(f), can also be easily understood by observing the impedance curve, Fig. 8(a), and the magnitude of the acoustic frequency components in Fig. 9(f). With the lowest impedance and the highest acoustic excitation amplitude, the amplitude of the first frequency component of the voltage output is highest in magnitude. Similar reasoning can explain the almost equal amplitude of the second and third frequencies in the voltage output. It is also seen that unlike the case study M2, the second, third and fourth frequency components of the voltage response are only an order lower as compared to the first component in M3, thus contributing to an increased voltage response than M2.

In previous case studies, the hybrid KZK-FE based model is used to predict the incident pressure field, which eliminates the computational cost of solving the nonlinear acoustic equation in the FE method. However, the structural response in the FE formulation is still solved in the time domain. This incurs a heavy computational cost in terms of memory and time. As a solution, it is proposed to solve the complete FE model in the frequency domain iteratively at each frequency of the nonlinear acoustic pressure field, such that the total voltage response, $V_{tot}$ is

$$V_{tot} = \sum_{n=1}^{4} V_n e^{i\omega_n t} \tag{24}$$

where $V_n$ is the amplitude of the voltage response at the $n^{th}$ acoustic excitation frequency, $\omega_n$. Using Eq. (24), the KZK-FE model in the frequency domain reduces the simulation time from hours/days to minutes. This reduction is because the time domain simulations need to be solved at each time step, which is inversely proportional to the frequency of the highest harmonic in the system, until a steady-state response is reached. Whereas in a complete frequency domain-based formulation, the simulations are solved only for pre-defined frequencies. However, this linear superposition, Eq. (24) is only possible when the natural frequencies of the structural modes of the disk are far apart, such that there is no mutual interference between the modes [57]. Also, the presence of any structural or geometrical nonlinearities can interfere with the neighboring frequency components of the excitation wave and make the Eq. (24) invalid. Notably, the summation of higher frequencies will contribute negligibly to the total output voltage when acoustic excitation frequencies do not coincide with the natural frequencies of the disk, such as in case M2. Thus, the summation is not needed for such cases. To validate the complete frequency domain formulation, the pressure field of M3 is used. The frequency-domain FE model is solved for each of the four acoustic wave frequency components separately, and the individual voltage responses are shown in Fig. 10. Since the frequency components of excitation waves are far apart, they excite only the intended resonant mode. The average power output in M4 is seen to be approximately equal to the power output of case study M3, 51.5 mW.

To summarize the four case studies, it is seen that as compared to a linear acoustic excitation, for the same given input power, a nonlinear excitation produces a higher power output in a UPT system. However, this increase in average power output is more significant when the frequencies of the higher overtones of the nonlinear acoustic wave coincide with the structural resonant frequencies of the disk. Consequently, not much increase in average power (6.2%) is observed between case studies M1 and M2. However, the average power increases significantly (75%) between case studies M2 and M3. It is to be noted that these values of the percentage increase in average output power between individual case studies may change when the constant damping ratio assumption is ignored, and separate damping ratios corresponding to each mode is employed. From these case studies, it can be concluded that accounting for acoustic nonlinearities that are either inherent in the system or generated intentionally, can lead to increased voltage response, as compared to assuming linear acoustic excitation. Moreover, case study M4 shows that the time domain nonlinear acoustic-structure interaction problem can be solved in a complete frequency domain and can reduce the computation time significantly.

Based on the conclusions drawn from the experimental and numerical investigations of the HIFU-UPT system, design recommendations for the three interwoven elements of the system are outlined. The first category consists of recommendations for the transmitter parameters. As seen in Eq. (2) and section 3.1, a change in effective geometrical parameters of the transmitter (aperture radius and radius-of-curvature) can change the dimensions of the focal zone as well as the gain in the amplitude of the focal pressure. Knowledge of the focal zone dimensions is crucial for the selection of the receiver size to obtain maximum power output. Moreover, an increase in input voltage to the transmitter increases the amplitude of the focal pressure nonlinearly (Fig. 5(a)), which may result in increased power output based on the discussion in section 3.4. The second category of design criteria is the acoustic parameters of the wave propagating medium. A change in medium parameters, $\beta$ and $\bar{A}$, affect the amplitude of the nonlinear pressure field at the focal point and consequently affect the output power response of the receiver. For example, to acoustically excite receivers placed inside the human body, the acoustic waves pass through multiple mediums with different values of acoustic parameters such as through fat ($\beta \approx 10$) and bones ($\bar{A} = 1.4$) [56]. As shown in Figs. 5(b) and (c), high values of these parameters affect the amplitude of the acoustic field at the receiver. Thus, modeling of acoustic nonlinearity due to wave kinematics or medium is essential to design a HIFU-UPT system. The final category for designing criteria is the receiver parameters. Since in a focused ultrasound field, the higher acoustic frequency components are harmonics, one possible scenario, in which they can actuate a receiver to yield significant power output, is when the structural resonance modes are harmonics of the fundamental mode. This is true for disks with a diameter-to-thickness ratio of less than 0.1 and greater than 20 [21, 36]. For such disks, the frequencies of the acoustic harmonics can coincide with the structural resonant modes of the disk to give a significant power output, as discussed in section 3.4. Besides, the receiver's power output can further be increased by adding matching layers and rectifier circuits, as proposed in [6].

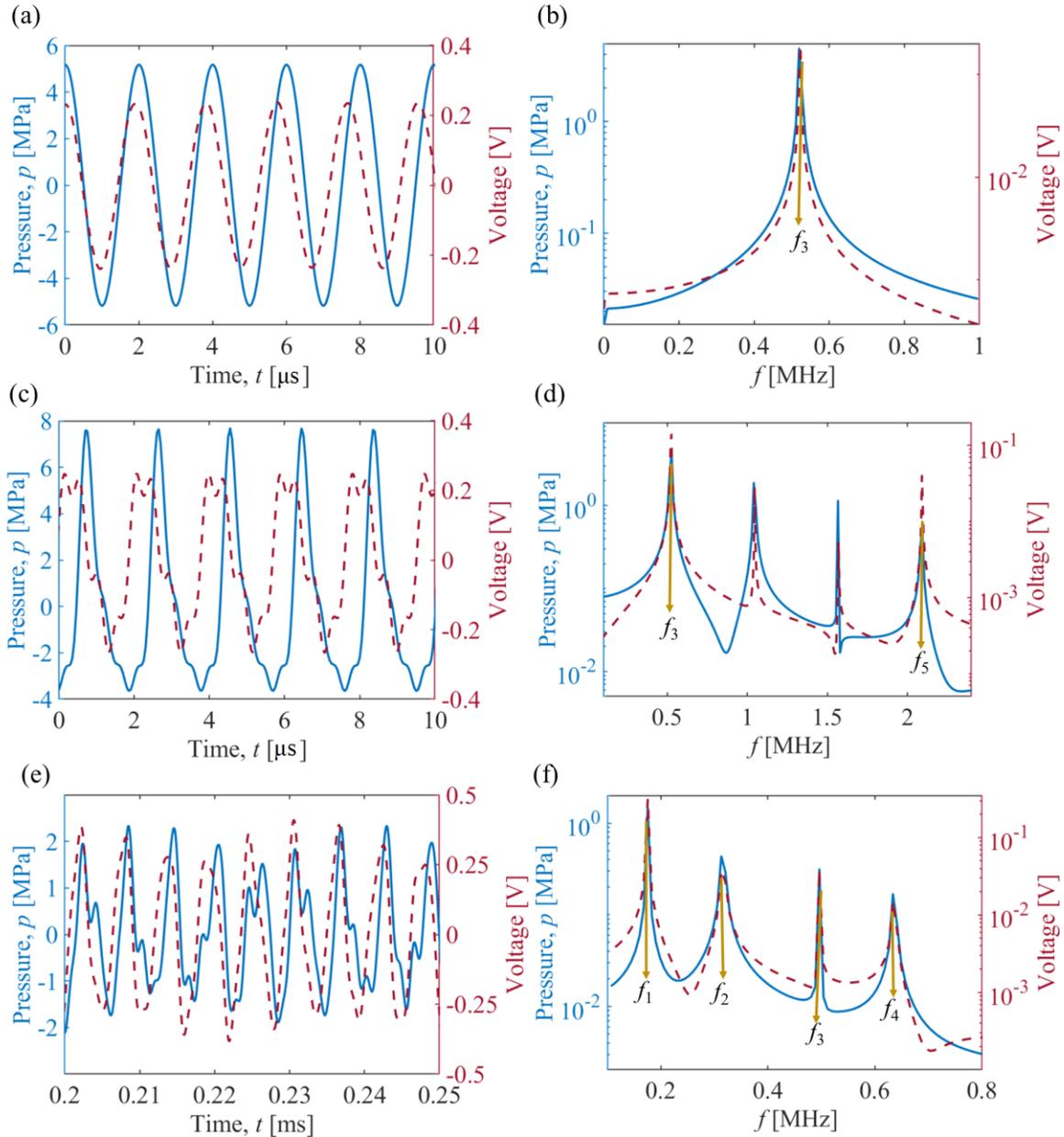

Fig. 9. The impinging pressure wave on the disk (solid blue line) and resulting voltage output (red dashed line) of the piezoelectric disk upon acoustic excitation in (a,c, and e) time domain and (b,d, and f) frequency domain, at $p_0 = 0.34$ MPa and $f = 0.5$ MHz. Case studies M1, M2 and M3 are shown by (a)-(b), (c)-(d), and (e)-(f) respectively. The arrows in the frequency domain plots point to the corresponding structural resonant frequencies in Fig. 8(a).

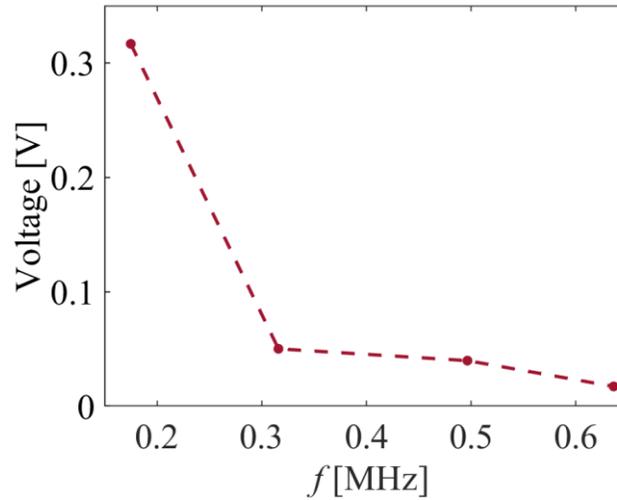

Fig. 10. Amplitudes of voltage output solved in the frequency domain for case study M4. The excitation frequencies and amplitudes of the incident pressure field are the same as the frequency components of the pressure field in case study M3.

## 4. Conclusions

We proposed a novel concept of using HIFU technology in UPT, for focusing the transmitted energy in space. This configuration strongly excites the receiver with dramatically reduced levels of energy input to the source in comparison to the case of unfocused power transmission, i.e., spherical and cylindrical sources. In a HIFU-UPT system, one of the dominant origins of nonlinearity is acoustic nonlinearity due to wave kinematics in HIFU fields. As the acoustic wave propagates towards the receiver, the wave distorts and becomes nonlinear with the generation of higher harmonics. In this work, the nonlinear acoustic field was calculated using the KZK equation for a focused source by taking into account the effects of diffraction, absorption, and nonlinearity in the medium. The pressure field obtained from the KZK model was used as an input to an FE-based acoustic-structure interaction formulation employed in COMSOL multiphysics.

The current studies for modeling UPT systems remain limited in capability due to three main assumptions. The assumptions are: (1) The piston-like deformation is assumed for the thickness mode. This assumption converts the two-dimensional axisymmetrical displacement of the disk to a one-dimensional one and significantly simplifies the model. However, this assumption fails for a finite-size (diameter-to-thickness ratio is greater than 0.1 and less than 20) transmitter or receiver, because the response of finite-size disks depends both on radial and thickness directions. (2) A linear acoustic field is assumed, which considers that all the energy is concentrated in the fundamental frequency component; thus, ignoring the effects of medium nonlinearity. (3) The acoustic-structure interaction effects on the receiver disk which include reflected, scattered, and blocked acoustic pressure distributions arising from the acoustic boundary conditions, are neglected. All these assumptions may lead to an inaccurate estimation of power output from the receiver. This study aimed to present a comprehensive model without taking these assumptions. The experimentally-validated FE multiphysics modeling approach aimed at filling a knowledge gap by considering the coupling of the acoustic nonlinear effects on the electroelastic response that lead to structural resonances of a finite-size piezoelectric disk receiver. It was assumed that no geometric or material nonlinearities were triggered in the piezoelectric disk receiver. However, the modeling and identification of electroelastic nonlinearities in UPT systems [68] can be combined with the proposed framework in this paper. The results showed that the existence of the HIFU high-level excitation can cause disproportionately large responses in the piezoelectric receiver if the frequency components in the nonlinear acoustic field coincide with the structural resonant frequencies of the receiver.

Moreover, by implementing the FE formulation for estimating the structural response of the receiver in the frequency domain instead of the time domain, the computational cost can be significantly reduced. This frequency-domain formulation was based on the linear superposition of voltage responses due to disk excitation at individual frequency components of the nonlinear acoustic field. The superposition is possible under the assumption that excitation frequencies are well separated to avoid mutual modal interference. The investigations of this work aimed to provide a guide for all those systems where piezoelectric disks of finite aspect ratios are operating in a nonlinear acoustic sound field.

**Acknowledgments**

This work was supported by the National Science Foundation (NSF), Grant No. ECCS-1711139, which is gratefully acknowledged. The authors would like to thank Marjan Bakhtiari-Nejad and Vamsi C. Meesala for helping with the experiments.

**Appendix**

$Y$, $e$ and, $\varepsilon$ are the $6\times 6$ elastic modulus at the constant electric field, $3\times 6$ piezoelectric coupling, and $3\times 3$ permittivity matrices for isotropic piezoelectric materials, given as

$$Y = \begin{pmatrix} Y_{11} & Y_{12} & Y_{13} & 0 & 0 & 0 \\ Y_{21} & Y_{22} & Y_{23} & 0 & 0 & 0 \\ Y_{31} & Y_{32} & Y_{33} & 0 & 0 & 0 \\ 0 & 0 & 0 & Y_{44} & 0 & 0 \\ 0 & 0 & 0 & 0 & Y_{55} & 0 \\ 0 & 0 & 0 & 0 & 0 & Y_{66} \end{pmatrix}, \; e = \begin{pmatrix} 0 & 0 & 0 & 0 & e_{15} & 0 \\ 0 & 0 & 0 & e_{24} & 0 & 0 \\ e_{31} & e_{32} & e_{33} & 0 & 0 & 0 \end{pmatrix} \text{ and}$$

$$\varepsilon = \begin{pmatrix} \varepsilon_{11} & 0 & 0 \\ 0 & \varepsilon_{22} & 0 \\ 0 & 0 & \varepsilon_{33} \end{pmatrix}$$

where $Y_{11} = Y_{22}$, $Y_{12} = Y_{21}$, $Y_{23} = Y_{32} = Y_{13} = Y_{31}$, $Y_{44} = Y_{55}$ and $Y_{66} = (Y_{11} - Y_{12})/2$. In the electrocoupling matrix, $e_{31} = e_{32}$ and $e_{24} = e_{15}$. For permittivity matrix, $\varepsilon_{11} = \varepsilon_{22}$ holds true.

# Acoustic-electroelastic modeling of piezoelectric disks in high-intensity focused ultrasound power transfer systems

**Aarushi Bhargava[a] and Shima Shahab[a, b,1]**

[a] Department of Biomedical Engineering and Mechanics, Virginia Polytechnic Institute and State University, Blacksburg, VA 24061, USA

[b] Department of Mechanical Engineering, Virginia Polytechnic Institute and State University, Blacksburg, VA 24061, USA

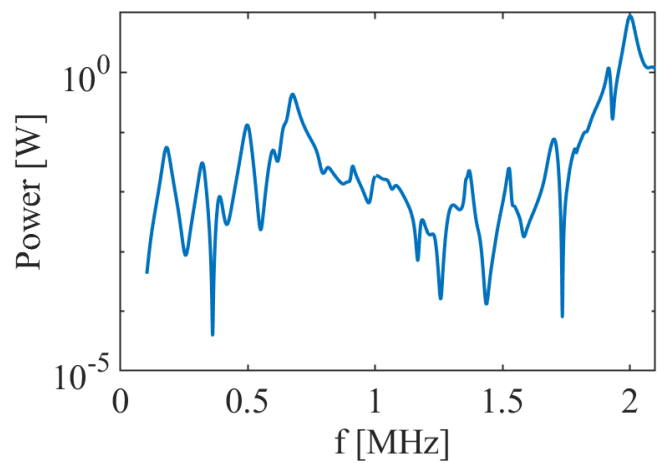

Figure S1. Normalized power output of the APC 760 disk on acoustic excitation. The incident pressure amplitude is equal for all frequencies and normalization is done with respect to maximum power.

---

[1] Corresponding author: sshahab@vt.edu

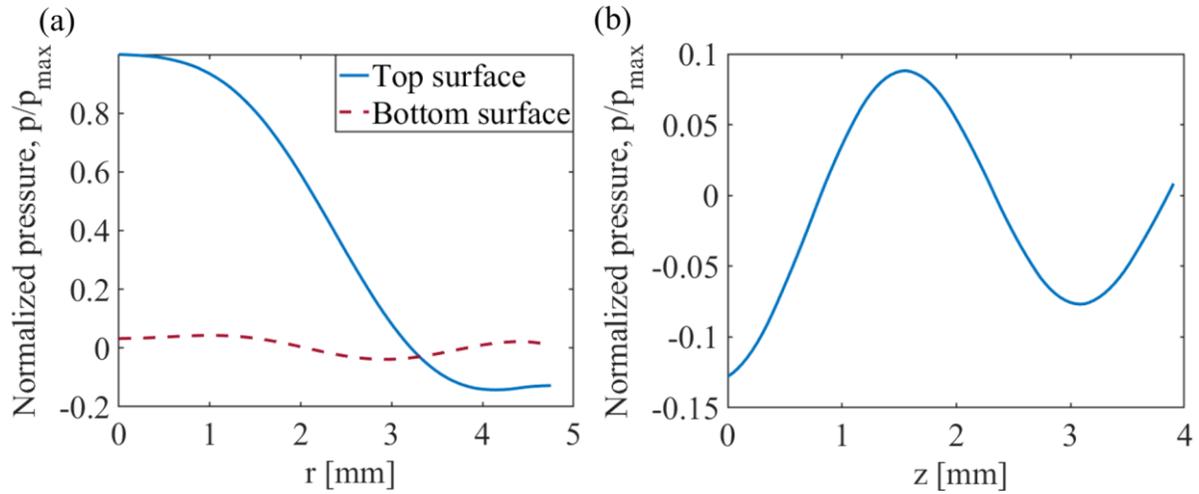

Figure S2. Normalized pressure profile at the (a) top and bottom radial surfaces and (b) circumferential surface of the disk. The normalization is done with respect to the maximum pressure from the three profiles.

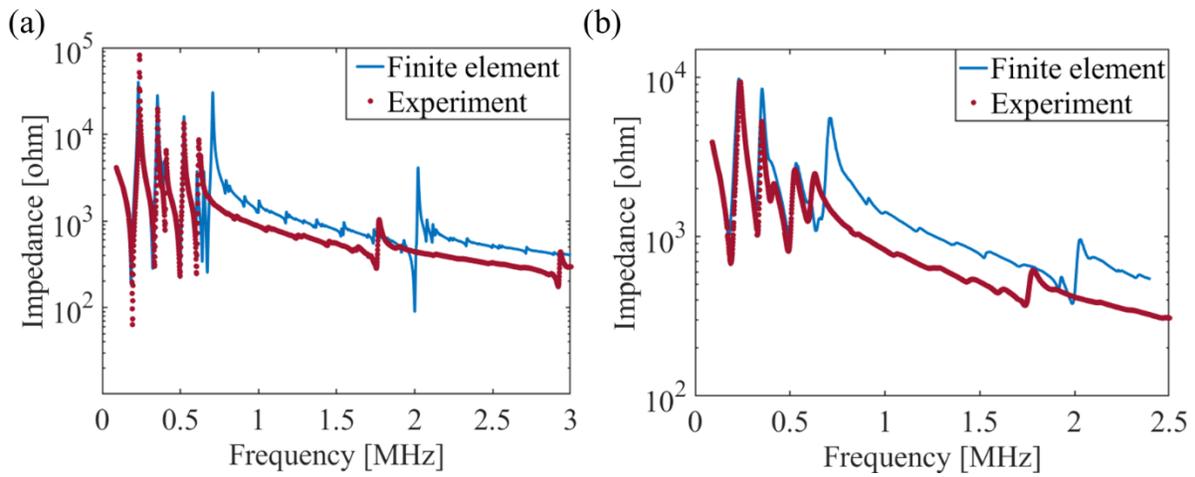

Figure S3. Electrical impedance curves obtained from finite element simulations (solid line) and experiments (dotted line) in (a) air and (b) water, for the APC760 disk.